\documentclass[aps,prdr,preprintnumbers,graphicx,twocolumn,
groupedaddress,showpacs]{revtex4}
\usepackage{amsmath,graphicx,graphics,epsf,psfrag}

\newcommand{\nl}{\nonumber\\}
\newcommand{\order}{{\cal O}}

\def\bm#1{\mbox{\boldmath$#1$\unboldmath}}

\begin{document}

\preprint{FERMILAB-PUB-06-182-T}

\title{Constraints on the form factors for $K\to \pi \ell\nu $ and 
implications for $|V_{us}|$}

\author{Richard J. Hill}
\affiliation{Fermi National Accelerator Laboratory, \\
P.~O.~Box 500, Batavia, IL 60510, U.S.A.}

\date{July, 2006}

\begin{abstract}
Rigorous bounds are established for the expansion coefficients governing  
the shape of semileptonic $K\to\pi$ form factors. 
The constraints enforced by experimental data from $\tau\to K\pi\nu$ 
eliminate uncertainties associated with model parameterizations 
in the determination of $|V_{us}|$. 
The results support the validity of a powerful expansion that 
can be applied to other semileptonic transitions.   
\end{abstract}


\maketitle

\section{Introduction \label{sec:intro} }

Semileptonic $K\to\pi\ell \nu$ ($K_{\ell 3}$) decays provide the most
precise determination of the Cabibbo-Kobayashi-Maskawa (CKM) matrix
element $|V_{us}|$~\cite{Blucher:2005dc}.  Presently, the dominant
uncertainty in the experimentally determined quantity $|V_{us}F_+(0)|$
arises from the unknown shape of the hadronic form factor $F_+(q^2)$,
as a function of momentum transfer
$q$~\cite{Alexopoulos:2004sy,Lai:2004kb,
Yushchenko:2004zs,Ambrosino:2006gn}.  The data can be fit to both
simplified pole models and series expansions, but the uncertainty
inherent in these simplifications is difficult to estimate, and the
difference between the resulting $|V_{us}F_+(0)|$ determinations
represents a systematic error.  Constraining this shape from first
principles and providing rigorous error estimates is an important
problem.

The remaining (and presently, dominant) error in the $K_{\ell 3}$
determination of $|V_{us}|$ arises from the normalization, $F_+(0)$.
$K\to\pi\ell\nu$ data (specifically $K_{\mu3}$) also constrain the
shape of the scalar form factor, $F_0(q^2)$.  This can potentially
reduce dominant errors in the theory normalization, either by avoiding
an extrapolation from zero recoil in lattice
determinations~\cite{Becirevic:2004ya}, or by fixing low-energy
constants of chiral perturbation theory~\cite{Bijnens:2003uy}.  At
present, the comparison between theory and experiment, and between
different experiments, is complicated by uncertainties in the form
factor parameterization.  Again, extracting as much information as
possible from the experimental data without model assumptions is an
important task.

This paper establishes rigorous bounds for the expansion coefficients
appearing in a general parameterization of the semileptonic $K\to\pi$
form factors.  The framework for the analysis is based on familiar
 arguments invoking analyticity and crossing
 symmetry~\cite{analyticity}.
However, in contrast to conventional arguments appealing to unitarity
and the evaluation of an operator product expansion (OPE), bounds for
the vector form factor are derived directly from experimental data for
hadronic tau decays, $\tau\to K\pi\nu$.  The result is more stringent
than can be obtained from the OPE analysis, and applies to a more
general class of parameterizations.\\[5pt]

\noindent{\it Analyticity and convergence. }
Form factors are defined as usual by the matrix element of the weak
vector current $V^\mu \equiv \bar{u}\gamma^\mu s$: ($q\equiv
p-p^\prime$, $\Delta_{K\pi}\equiv m_K^2-m_\pi^2$)
\begin{eqnarray}
&&\langle \pi^+(p^\prime)|V^\mu|\bar{K}^0(p)\rangle \\ 
&& = 
F_+(q^2)\left( p^\mu+p^{\prime\mu} 
- {\Delta_{K\pi}\over q^2}q^\mu\right) \nonumber 
+  F_0(q^2){\Delta_{K\pi}\over q^2} q^\mu  \,. 
\end{eqnarray}
$F_+(t=q^2)$ and $F_0(t)$ can be extended to analytic functions
throughout the complex $t$ plane, except along a branch cut on the
positive real axis starting at $K\pi$ production threshold $t=t_+$
[$t_\pm \equiv (m_K\pm m_\pi)^2$].  The cut plane is mapped onto the
unit circle by
\begin{equation}\label{eq:zdef}
t \to z(t,t_0) \equiv { \sqrt{t_+ - t} - \sqrt{t_+ - t_0} 
\over \sqrt{t_+ - t} + \sqrt{t_+ - t_0}} \,,
\end{equation}
where $t_0 \in (-\infty, t_+)$ is the point mapping onto $z=0$.~%
\footnote{ 
  By the Riemann mapping theorem, this conformal
  transformation is unique up to the choice of $t_0$ and an overall
  phase See e.g. \cite{ahlfors}.  }  
The form factors are analytic in
$|z|<1$, and so may be expanded in a convergent power series:
\begin{equation}
\label{eq:Fexpand}
F(t)= {1\over \phi(t,t_0,Q^2)} \sum_{k=0}^\infty a_k(t_0,Q^2)\, z(t,t_0)^k \,,
\end{equation}
where $\phi$ is an as-yet arbitrary function analytic in $|z|<1$,
which may depend on one or more parameters, denoted generically (and
for reasons to become clear) by $Q^2$.

The function $z(t,t_0)$ sums an infinite number of terms, transforming
the original series, naively an expansion involving $t/t_+ \lesssim
0.3$, into a series with a much smaller expansion parameter.  For
example, the choice $t_0=t_+(1-\sqrt{1-t_-/t_+})$ minimizes the
maximum value of $z$ occurring in the semileptonic region, and for
this choice $|z(t,t_0)| \lesssim 0.047$.  The function $\phi$ and the
number $t_0$ may be regarded as defining a ``scheme'' for the
expansion.  The expansion parameter $z$ and coefficients $a_k$ are
then ``scheme-dependent'' quantities, with the scheme dependence
dropping out in physical observables such as $F(t)$.

Neglecting terms beyond $z^N$ in (\ref{eq:Fexpand}) introduces a
relative normalization error $\Delta F /F = \order( z^{N+1} )$.
Similarly, the error on the relative slope involves terms of order
$(N+1)z^N$.  Since $|z|^2 \lesssim 2\times 10^{-3}$ and $|z|^3
\lesssim 1\times 10^{-4}$, simple power counting in $z$ yields strong
constraints on the impact of higher-order terms in the expansion,
provided that the coefficients $a_k/a_0$ are well-behaved.\\[5pt]

\noindent{\it Crossing symmetry and form factor bounds.}
It is important in practice to determine 
whether large ``order unity'' coefficients $a_k/a_0$ could upset the 
formal power counting in $z$. 
To address this question, a norm may be defined via 
\begin{eqnarray}\label{eq:norm}
|| F ||^2 \equiv \sum_{k=0}^\infty a_k^2 
&=& {1\over 2\pi i} \oint {dz\over z} |\phi F|^2  \nl
&=& {1\over \pi} \int_{t_+}^\infty {dt \over t - t_0 } 
\sqrt{ t_+ - t_0\over t-t_+ } |\phi F|^2 \,. 
\end{eqnarray} 
By crossing symmetry, 
the norm can be evaluated 
using form factors for the related process of $K\pi$ production. 
The following sections investigate 
bounds on the integral appearing on the right hand side of (\ref{eq:norm}). 

\section{Vector Form Factor Constraints \label{sec:vector}}

\begin{figure}
  \psfrag{2}{$Q$}
  \epsfxsize=7.7cm \vspace{-0.3cm} \centerline{\epsffile{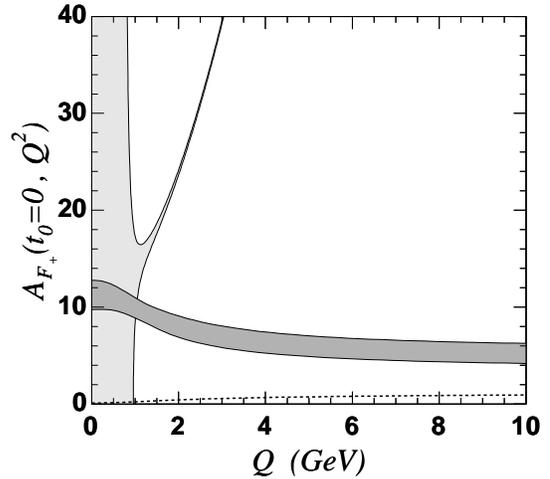}}
  \vspace{-0.5cm}
\caption{\label{fig:vector}Bounds on the expansion coefficients for the 
  vector form factor.  The top (light) band represents the unitarity
  bound, and the lower (dark) band is a direct evaluation from $\tau$
  decay and perturbative QCD.  The perturbative contribution is shown
  separately as the dashed line.  }
\end{figure}

To compare with unitarity predictions, and to motivate a choice of
$\phi$ in (\ref{eq:Fexpand}), we consider the correlation function,
\begin{eqnarray}\label{eq:ope}
\Pi^{\mu\nu}(q) &\equiv& i\int\! d^4x\, e^{iq\cdot x} \langle 0 | T\left\{ V^\mu(x) V^{\nu}(0)^\dagger \right\} | 0 \rangle  \nl
&=& ( - g^{\mu\nu}q^2+q^\mu q^\nu) \Pi_1(q^2) + q^\mu q^\nu \Pi_0(q^2) \,. 
\end{eqnarray} 
An unsubtracted dispersion relation can be written for the quantity:
($Q^2=-q^2$)
\begin{equation}
\chi_1(Q^2) \equiv {1\over 2} {\partial^2 \over \partial (q^2)^2}\left[ q^2\Pi_1 \right]  
= {1\over \pi} \int_0^\infty\! dt\,{ t {\rm Im}\Pi_1(t) \over (t+Q^2)^3 } \,.  
\end{equation}
Assuming isospin symmetry, the contribution of all $K\pi$ states to
the (positive) spectral function ${\rm Im}\Pi_1(t)$ is
\begin{equation}\label{eq:Pi1}
 {3\over 2}{1\over 48\pi}{ [(t-t_+)(t-t_-)]^{3/2}\over t^3} 
|F_+(t)|^2\theta(t-t_+) \le {\rm Im}\Pi_1(t) \,.
\end{equation} 
Choosing: [note that $|z|=1$ along the contour in (\ref{eq:norm}) ]
\begin{eqnarray}\label{eq:phiplus}
&&\phi_{F_+}(t,t_0,Q^2) = \sqrt{1\over 32\pi} {z(t,0)\over -t}
\left(z(t,-Q^2)\over -Q^2-t\right)^{3/2} \nl
&& \times
\left(z(t,t_0)\over t_0-t\right)^{-1/2} 
\left(z(t,t_-)\over t_--t\right)^{-3/4} 
{t_+-t\over (t_+-t_0)^{1/4}} \,,
\end{eqnarray}
then yields the inequality:~%
\footnote{ The quantity $A$ defined in \cite{Becher:2005bg} is
  normalized to the leading OPE prediction: $A \equiv (32\pi^2
  m_b^2/3) a_0^2 A_{F_+}^2$.  } 
[note that
$a_0(t_0,Q^2)=\phi(t_0,t_0,Q^2)F(t_0)$]
\begin{equation}\label{eq:Aplusdef}
A_{F_+}^2(t_0,Q^2) \equiv \sum_{k=0}^\infty {a_k^2\over a_0^2}
\le {\chi_1(Q^2)\over |\phi_{F_+}(t_0,t_0,Q^2) F_+(t_0)|^2} \,. 
\end{equation} 
For $Q \gg \Lambda_{\rm QCD}$, $\chi_1(Q^2)$ can be reliably
calculated using the OPE in (\ref{eq:ope}).  Collecting results from
the literature~\cite{Gorishnii:1990vf,Generalis:1983hb},
we have at renormalization scale $\mu=Q$:~%
\footnote{ Results are for $n_f=3$ light flavors, in the modified
  minimal subtraction scheme.  }
\begin{eqnarray}\label{eq:chi1}
\chi_1(Q^2) 
&=& {1\over 8\pi^2 Q^2}\bigg\{ 
1+ {\alpha_s\over\pi} -0.062 \alpha_s^2 
- 0.162\alpha_s^3 + \dots \nl
&& 
+{1\over Q^2}\bigg[ 
 -{3\over 2}{m_s^2} + \dots \bigg]  \\
&&
+{8\pi^2\over Q^4}\bigg[ -m_s\langle \bar{u}u \rangle 
- {\alpha_s\over 12\pi } \langle G^2\rangle 
+ \dots \bigg] + \dots  \bigg\} \,, \nonumber
\end{eqnarray}
where corrections of order $m_u/m_s$ are neglected, and  
$m_s(2\,{\rm GeV})=0.087(8)\,{\rm GeV}$~\cite{Mason:2005bj},  
$-m_s\langle \bar{u}u \rangle - \alpha_s \langle G^2\rangle /12\pi 
= -0.0001(8) {\rm GeV}^4$~\cite{Narison:2002hk}.   
The light band in Fig.~\ref{fig:vector} shows the 
resulting bound on the quantity $A_{F_+}(t_0,Q^2)$, setting $t_0=0$ 
and $F_+(0) \approx 1$.
The perturbative uncertainty is estimated by 
varying $\mu^2$ from $Q^2/2$ to $2Q^2$, and allowing for higher-order 
contributions of relative size $\pm 1\times \alpha_s^4$. 
Uncertainties from perturbative and power corrections are 
small above $Q= 2\,{\rm GeV}$, but become significant below this scale.   
The width of the band represents a $1\sigma$ contour 
obtained by adding uncertainties in 
perturbative and power corrections linearly. 

\begin{figure}
  \psfrag{2}{$Q$} \epsfxsize=7.7cm \vspace{-0.3cm}
  \centerline{\epsffile{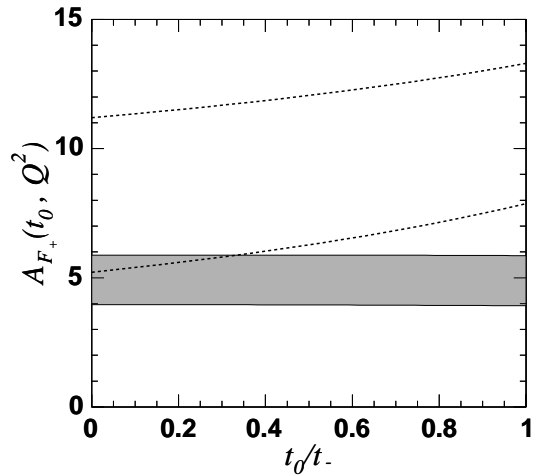}} \vspace{-0.5cm}
\caption{\label{fig:t0vector}
  Bounds on $A_{F_+}$ in (\ref{eq:Aplusdef}), as a function of $t_0$.
  The dashed lines are for the default choice of $\phi_{F_+}$ in
  (\ref{eq:phiplus}), with $Q=0$ (top) and $Q=10\,{\rm GeV}$ (bottom).
  The solid band gives the ($Q$-independent) result when
  $\phi_{F_+}\equiv 1$.  }
\end{figure}

Although the OPE breaks down at small $Q$, the norm in (\ref{eq:norm})
remains perfectly well defined.  The dark band in
Fig.~\ref{fig:vector} shows the result of evaluating the integral in
(\ref{eq:norm}) using $\tau\to K\pi\nu$ decay
data~\cite{Barate:1999hj} for the region $t_+ < t < m_\tau^2$, and
estimating the remaining contribution from $t> m_\tau^2$ using
perturbative QCD.  Uncertainties from the $\tau$ decay data are
conservatively estimated by taking the maximum (minimum) value of the
weight function in (\ref{eq:norm}) for each bin, and multiplying by
the corresponding upper (lower) bound of the $1\sigma$ interval for
the $K\pi$ component of the spectral function measured in
\cite{Barate:1999hj}.  It may be noted that for the present purpose,
there is no need to resolve the underlying resonance structure of the
spectral function, nor to subtract the (small) $S$-wave contribution.

The remaining perturbative contribution is estimated using $F_+(t)
\sim 8\pi f_\pi f_K \alpha_s/ t$~\cite{Lepage:1980fj}, conservatively
setting $\alpha_s\approx 1$, and assigning $100\%$ uncertainty to the
result.  The effect is $\Delta A_{F_+}^2 \lesssim 2$ for all values of
$Q\in (0,10\,{\rm GeV})$, and is insignificant when added in
quadrature to the much larger contribution from the resonance region.

As illustrated by Fig.~\ref{fig:t0vector}, the dependence of $A_{F_+}$
on $t_0$ is very mild.  Here we again set $F_+(t_0)\approx 1$; this is
strictly an overestimate, and accounts for much of the $t_0$ variation
in the figure.  The norm for all schemes (\ref{eq:phiplus}) with
$0<t_0<t_-$ and $0<Q<10\,{\rm GeV}$ satisfies
\begin{equation}\label{eq:vbound}
A_{F_+}(t_0,Q^2) \lesssim 13 \,. 
\end{equation}
More refined values for $A_{F_+}$ can be obtained from
Figs.~\ref{fig:vector} and \ref{fig:t0vector} 
at particular values of $t_0$ and $Q$.

The result from the OPE represents an {\it upper bound} for $A_{F_+}$.
For large $Q$, the leading contributions to $A_{F_+}$ are
$\order(Q^0)$, coming from the regions $t\sim\Lambda_{\rm QCD}^2$, and
$t\sim Q^2$
in (\ref{eq:norm}).~%
\footnote{ Counting $m_{\pi,K}^2 \sim \hat{m}\Lambda_{\rm QCD}$ gives
  $A_{F_+} \sim \Lambda_{\rm QCD}/\hat{m}$.  Since $m_K/\Lambda_{\rm
    QCD}$ is not very small, this does not lead to a large
  enhancement.  For $Q^2\lesssim \hat{m}\Lambda_{\rm QCD}$ the result
  is simply $A_{F_+} \sim 1$, with dominant contributions from $t\sim
  \hat{m}\Lambda_{\rm QCD}$.  } 
The OPE overestimates $A_{F_+}$ by a
factor $\sim (Q/\Lambda_{\rm QCD})^2$.  As Fig.~\ref{fig:vector}
shows, although the OPE evaluation becomes exceedingly precise, the
resulting unitarity bound begins to wildly overestimate the norm of
the form factor already at $Q\sim 2-3\,{\rm GeV}$.

\section{Scalar Form Factor Constraints \label{sec:scalar}}

\begin{figure}
  \epsfxsize=7.7cm \vspace{-0.3cm} \centerline{\epsffile{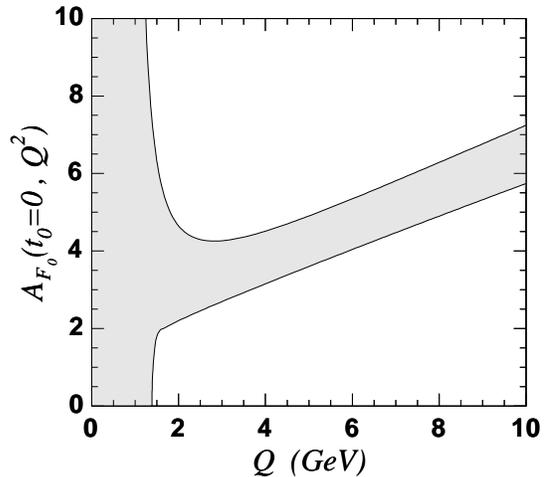}}
  \vspace{-0.5cm}
\caption{\label{fig:scalar} Bounds on the expansion coefficients for the scalar
  form factor.  }
\end{figure}

In the scalar case, an unsubtracted dispersion relation can be written
for the quantity
\begin{equation}
\chi_0(Q^2)\equiv {\partial \over \partial q^2} \left[ q^2\Pi_0 \right] 
= {1\over \pi}\int_0^\infty\!dt\, {t {\rm Im}\Pi_0(t) \over (t+Q^2)^2} \,. 
\end{equation}
Noticing that
\begin{equation}
{\rm Im} \Pi_0(t) \ge {3\over 2} {t_+ t_- \over 16\pi} 
{[(t-t_+)(t-t_-)]^{1/2}\over t^3} |F_0(t)|^2 \theta(t-t_+) \,,  
\end{equation}
and defining
\begin{eqnarray}\label{eq:phizero}
\phi_{F_0}(t,t_0,Q^2) &=& 
\sqrt{3t_+ t_-\over 32\pi} {z(t,0)\over -t} 
{z(t,-Q^2)\over -Q^2-t} 
\left(z(t,t_0)\over t_0-t \right)^{-1/2} \nl 
&& \times 
\left(z(t,t_-)\over t_--t \right)^{-1/4}
{\sqrt{t_+ - t} \over (t_+ - t_0)^{1/4}}  \,, 
\end{eqnarray}
yields the inequality
\begin{equation}\label{eq:Azerodef}
A_{F_0}^2(t_0, Q^2) \equiv \sum_k {a_k^2\over a_0^2} 
\le {\chi_0(Q^2) \over |\phi_{F_0}(t_0,t_0,Q^2) F_0(t_0)|^2 }
\,. 
\end{equation} 
Again, $\chi_0(Q^2)$ can be reliably calculated when 
$Q\gg\Lambda_{\rm  QCD}$~\cite{Chetyrkin:1996sr,Generalis:1983hb}:
\begin{eqnarray}
\chi_0(Q^2) &=& 
{3m_s^2\over 8\pi^2 Q^2} 
\big[ 1 + 1.80\alpha_s 
+ 4.65\alpha_s^2 + 15.0\alpha_s^3 + \dots \big] \nl 
&& + {1 \over Q^4}\big[m_s\left( \langle\bar{u} u\rangle - \langle \bar{s} s \rangle \right) + \dots
 \big] + \dots  \,,
\end{eqnarray}  
where $m_s(2\,{\rm GeV}) = 0.087(8) \, {\rm GeV}$~\cite{Mason:2005bj},
$m_s(\langle\bar{u} u\rangle - \langle \bar{s} s \rangle) =
-0.0006(3)\,{\rm GeV}^4$~\cite{Narison:2002hk}.  The light band in
Fig.~\ref{fig:scalar} shows the $1\sigma$ contour obtained by 
adding errors from all sources linearly, as described after (\ref{eq:chi1}); 
in this case, higher-order contributions of relative size $\pm 10\times \alpha_s^4$ 
are included.~%
\footnote{ At $Q^2=2\,{\rm GeV}^2$, the result for $\chi_0$ is consistent with a
  related study in \cite{Bourrely:2005hp}, where however uncertainties
  due to the poor convergence of the perturbation series, and the
  effects of power corrections were not considered.  }

The behavior of the unitarity bound with respect to $Q$ is milder than in the
vector case.  For large $Q$, the leading contributions to $A_{F_0}$ 
are $\order(Q^0)$, coming from the regions 
$t\sim\Lambda_{\rm QCD}^2$ and $t\sim Q^2$ in (\ref{eq:norm}).~%
\footnote{ For $Q^2\lesssim \hat{m}\Lambda_{\rm QCD}$, $A_{F_0} \sim
  1$, with dominant contributions from $t\sim \hat{m}\Lambda_{\rm
    QCD}$.  }
Again, it should be remembered that the OPE gives an
{\it upper bound} to $A_{F_0}$, although for this case the bound is
overestimated by only a single power of $Q/\Lambda_{\rm QCD}$.  While it
is not possible to rigorously bound the norm of $F_0$ using the OPE
for schemes with $Q\lesssim \Lambda_{\rm QCD}$, there is no parametric
enhancement of the integral in (\ref{eq:Azerodef}) for these $Q$
values.  A conservative bound on the norm for all schemes with
$0<t_0<t_-$ and $0<Q<10\,{\rm GeV}$ is
\begin{equation}\label{eq:sbound}
A_{F_0}(t_0,Q^2) \lesssim 5 \,. 
\end{equation}
In principle, this bound could be improved by extracting the 
$K\pi$ component of the scalar spectral function from $\tau\to K\pi\nu$. 

\section{Implications for $K_{\ell 3}$ and $|V_{us}|$}

To implement the expansion (\ref{eq:Fexpand}) we must choose a
``scheme'' specified by the functional form of $\phi$, and by the
values of $Q^2$ and $t_0$.  Possible candidates for $\phi$ are
$\phi=1$, or $\phi$ given by (\ref{eq:phiplus}) and
(\ref{eq:phizero}).  The latter choice ensures that the norm in
(\ref{eq:norm}) is independent of $t_0$ and relates in a simple way to
the spectral functions extracted from $\tau$ decay; with, say
$Q^2=2\,{\rm GeV}^2$, this choice also allows comparison to OPE bounds.~%
\footnote{ For heavy-quark systems the choice of $\phi$ in
  (\ref{eq:phiplus}) and (\ref{eq:phizero}) also ensures that $A_{F}$
  does not scale as some power of the heavy-quark
  mass~\cite{Becher:2005bg}.  }

The value of $t_0$ can be selected according to various criteria.  For
example, the choice $t_0=t_+(1-\sqrt{1-t_-/t_+})$ minimizes the
maximum value of $z$ throughout the semileptonic range, while $t_0=0$
simplifies the translation between coefficients $a_k$ and the
derivatives of the form factor at $t=0$.  Alternatively, the choice of
$t_0$ can be used to minimize correlations between shape parameters.

When only a single form factor contributes (as in $K_{e3}$ decays),
the error correlations between expansion coefficients $a_k(t_0,Q^2)$
measured in an experiment with near ideal acceptance and resolution
are completely determined without a priori knowledge of the form
factor shape.  With $\Delta\Gamma_i/\Gamma_i \propto
1/\sqrt{\Gamma_i}$ where $\Gamma_i \sim p(t_i)^3 F_+(t_i)^2$ are
partial rates measured in intervals of $t$, it follows that the error
matrix $U$ determined from a $\chi^2$ fit is: (here $p=|\bm{p}|$ is
the pion momentum in the kaon rest frame)
\begin{equation}
U^{-1}_{kk^\prime} \equiv 
{1\over 2}{\partial^2\chi\over \partial a_k \partial a_{k^\prime}} 
\propto
\int_0^{t_-}\! dt\, {p(t)^{3} \over \phi_{F_+}(t,t_0,Q^2)^2}
z(t,t_0)^{k+k^\prime} \,. 
\end{equation}
If the $K_{e3}$ data is fit to a normalization proportional to
$F_+(0)$ and two shape parameters $a_1(t_0,Q^2)/a_0(t_0,Q^2)$ and
$a_2(t_0,Q^2)/a_0(t_0,Q^2)$, the correlation between $a_1/a_0$ and
$a_2/a_0$ vanishes in the ideal case for $t_0\approx 0.39\,t_-$
(with $\phi_{F_+}$ as in (\ref{eq:phiplus}) and $Q^2=2\,\rm GeV^2$).~%
\footnote{ This value of $t_0$ is insensitive to the precise form
  factor shape, e.g.  whether $F_+$ is assumed constant, or is input
  from experimental data.  Since Ref.~\cite{Alexopoulos:2004sy}
  measures a distribution in transverse momentum, $t_\perp^\pi =
  m_K^2+m_\pi^2-2m_K\sqrt{m_\pi^2+\bm{p}_\perp^2}$, the partial rates
  $d\Gamma/dt_\perp^\pi$ involve a smearing over different $t$ values.
  Taking this into account, the correlations remain very insensitive
  to the form factor shape, and are minimized in the ideal case for
  $t_0 \approx 0.37\,t_-$ (at $Q^2=2\,{\rm GeV^2} )$.  }
 
\subsection{$F_+$ and $K_{e3}$} 

Only the vector form factor is relevant for the electron mode.  From
(\ref{eq:vbound}) and $|z| \lesssim 0.05$ [at
$t_0=t_+(1-\sqrt{1-t_-/t_+})$], the quadratic term, $a_2 z^2$ in
(\ref{eq:Fexpand}), can affect the normalization by an amount not
greater than $A_{F_+} |z|^2 < 0.029$.  To establish a precision
significantly smaller than this, $a_2$ must be constrained by data.
The cubic term is bounded by $A_{F_+} |z|^3 < 0.0014$.  These errors
can also be robustly estimated by including higher order terms in fits
to the experimental data, under the constraint (\ref{eq:vbound}).
Extraction of other observables, such as the form factor slope and
curvature at $t=0$ can be treated similarly.

The simple Taylor expansion in $t$ converges much more slowly than the
$z$ expansion in (\ref{eq:Fexpand}), and it is difficult to reliably
bound the errors associated with higher-order terms in fits to the
truncated $t$ series.  The systematic approach advocated here should
remedy this situation, and make a more definitive
comparison between different experiments possible.~%
\footnote{ Another commonly used parameterization, the pole model,
  suffers from similar difficulties.  To avoid biases, this model
  would need to be generalized.  A pedestrian, but rigorous, approach
  is to break apart the dispersive integral into the sum of effective
  poles~\cite{Hill:2005ju,Becher:2005bg}.  Although there is no
  analogue of the OPE here, we can establish a bound on the
  coefficients of the effective poles using the $\tau$-decay data in
  \cite{Barate:1999hj} (and a negligible perturbative contribution) to
  obtain $(1/\pi)\int_{t_+}^\infty\!dt^\prime\,
  |F_+(t^\prime)|/t^\prime \lesssim 2.3\pm 0.4$.  }

\subsection{$F_0$ and $K_{\mu3}$}

The scalar form factor can be probed in the muon mode.  With enough
precision, constraining the slope of this form factor could aid
lattice and chiral perturbation theory estimates of $F_+(0)=F_0(0)$ by
eliminating extrapolation errors~\cite{Becirevic:2004ya}, or by
determining higher-order constants in the chiral
Lagrangian~\cite{Bijnens:2003uy}.  In a general scheme: (the prime
denotes derivative with respect to $t$)
\begin{equation}
{F_0^\prime(0)\over F_0(0)}\! =\! - 
{\phi_{F_0}^\prime(0,t_0,Q^2)\over \phi_{F_0}(0,t_0,Q^2)} 
+ z^\prime(0,t_0){ {a_1\over a_0} + {2a_2\over a_0} z(0,t_0) 
+ \dots \over 1 + {a_1\over a_0} z(0,t_0) + \dots } \,.
\end{equation}
Present experiments do not strongly constrain the coefficients beyond
the linear $a_1 z$ term in (\ref{eq:Fexpand}).  From (\ref{eq:sbound})
and $|z| \lesssim 0.05$
it is straightforward to estimate the resulting uncertainty on the
slope: $\Delta [m_\pi^2 F_0^\prime(0)/F_0(0) ] \approx A_{F_0} \times
0.001 < 0.005$.  This error can also be robustly estimated by
including higher order terms in fits to the experimental data, under
the constraint (\ref{eq:sbound}).

To further constrain the slope requires either more precise
semileptonic data, or independent constraints on the form factor.  One
type of constraint appeals to symmetry principles, in particular at
the Callan-Treiman point where $F_0(\Delta_{K\pi}) = {f_K/f_\pi} +
\Delta_{CT}$ with a controllably small correction,
$\Delta_{CT}$~\cite{Gasser:1984ux}.  Unfortunately, this point is too
far removed from $t=0$ to have a very strong influence on
the extraction of $F^\prime_0(0)/F_0(0)$.~%
\footnote{ Semileptonic data combined with the bound (\ref{eq:sbound})
  yields a more precise value for the slope than is obtained from
  combining unitarity with the Callan-Treiman point in the absence of
  data, cf. \cite{Bourrely:2005hp}.}  
As a second type of constraint,
the phase of the form factor above $K\pi$ threshold can be determined
from $K\pi$ scattering.  Even supposing arbitrarily precise scattering
data were available up to the inelastic threshold, $t_{\rm inel} =
(m_K+3m_\pi)^2$, it is easy to see that dramatic constraints are not
expected in the semileptonic region.  Since the additional data
relates to the form factor at points along the cut in the $t$-plane,
where $|z(t,t_0)|=1$, constraints from the scattering data can be
satisfied by tuning higher-order coefficients $a_k$ in
(\ref{eq:Fexpand}), which have relatively minor impact in the
semileptonic region.  More quantitatively, with arbitrarily precise
data, a suitable phase redefinition of the form factor (a particular
choice of $\phi$) can postpone the branch point in $\phi F$ until
$t_{\rm inel}$.  Replacing $t_+$ with $t_{\rm inel}$ in
(\ref{eq:zdef}) , we see that the maximum size of the effective
expansion parameter can then be made as small as $|z| \lesssim 0.021$
throughout the semileptonic region, compared to $|z| \lesssim 0.047$
when the branch point appears at $t_+$.  Thus while some improvement
is possible by directly incorporating the experimental scattering
data, more dramatic gains in precision require further dynamical
inputs or assumptions~\cite{Jamin:2006tj}.

\section{Implications beyond $K_{\ell 3}$}

The general parameterization (\ref{eq:Fexpand}), with constraints
(\ref{eq:vbound}) and (\ref{eq:sbound}), can be used to systematically
analyze the $K\to\pi\ell\nu$ data without model assumptions.  This
should eliminate a dominant uncertainty in the experimental
determination of $|V_{us}F_+(0)|$, and allow for a more definitive
comparison of shape observables measured by different experiments.

$K_{\ell 3}$ decays provide a unique 
opportunity to investigate the convergence properties 
of the expansion (\ref{eq:Fexpand}).  Precision data can be
used to directly constrain the first few coefficients, and 
the existence of a heavy lepton, $m_\tau \gg m_K+m_\pi$, 
makes it possible to establish bounds on the dominant vector 
form factor that are both more stringent and more generally applicable 
than those obtained from an OPE analysis. 
This provides a direct test of scaling arguments that apply 
also in the charm and bottom systems~\cite{Becher:2005bg}.

\vspace{3mm}
\noindent {\it Acknowledgements.} 
The author acknowledges a number of fruitful discussions with 
A.~Glazov and R.~Kessler.  Thanks also to T.~Becher for comments on the 
manuscript, and M.~Davier for a useful remark 
concerning the details of Ref.~\cite{Barate:1999hj}.   
Fermilab is operated by Universities Research Association Inc. 
under contract DE-AC02-76CH03000 with the U.S. Department of Energy.


\begin{thebibliography}{99}

\bibitem{Blucher:2005dc}
  For a review see: E.~Blucher {\it et al.},
  hep-ph/0512039.

\bibitem{Alexopoulos:2004sy}
  T.~Alexopoulos {\it et al.}  [KTeV Collaboration],
  Phys.\ Rev.\ D {\bf 70}, 092007 (2004).

\bibitem{Lai:2004kb}
  A.~Lai {\it et al.}  [NA48 Collaboration],
  Phys.\ Lett.\ B {\bf 604}, 1 (2004).

\bibitem{Yushchenko:2004zs}
  O.~P.~Yushchenko {\it et al.},
  Phys.\ Lett.\ B {\bf 589}, 111 (2004).

\bibitem{Ambrosino:2006gn}
  F.~Ambrosino {\it et al.}  [KLOE Collaboration],
  Phys.\ Lett.\ B {\bf 636}, 166 (2006).

\bibitem{Becirevic:2004ya}
  D.~Becirevic {\it et al.},
  Nucl.\ Phys.\ B {\bf 705}, 339 (2005).

\bibitem{Bijnens:2003uy}
  J.~Bijnens and P.~Talavera,
  Nucl.\ Phys.\ B {\bf 669}, 341 (2003).
  For further references, see: J.~Bijnens,
  hep-ph/0604043.

\bibitem{analyticity}  
  See e.g.: 
  C.~Bourrely, B.~Machet and E.~de Rafael,
  Nucl.\ Phys.\ B {\bf 189}, 157 (1981), and references therein. 
%
  Early applications to heavy-meson decays and related developments can be found in:
  C.~G.~Boyd, B.~Grinstein and R.~F.~Lebed,
  Phys.\ Rev.\ Lett.\  {\bf 74}, 4603 (1995);
%
  L.~Lellouch,
  Nucl.\ Phys.\ B {\bf 479}, 353 (1996);
%
  I.~Caprini and M.~Neubert,
  Phys.\ Lett.\ B {\bf 380}, 376 (1996).

\bibitem{ahlfors}
  L.~V.~Ahlfors, 
  {\it Complex Analysis,} McGraw-Hill, 1953. 

\bibitem{Becher:2005bg}
  T.~Becher and R.~J.~Hill,
  Phys.\ Lett.\ B {\bf 633}, 61 (2006).

\bibitem{Gorishnii:1990vf}
  S.~G.~Gorishnii, A.~L.~Kataev and S.~A.~Larin,
  Phys.\ Lett.\ B {\bf 259}, 144 (1991).
  L.~R.~Surguladze and M.~A.~Samuel,
  Phys.\ Rev.\ Lett.\  {\bf 66}, 560 (1991)
  [Erratum-ibid.\  {\bf 66}, 2416 (1991)].
  K.~G.~Chetyrkin,
  Phys.\ Lett.\ B {\bf 391}, 402 (1997). 

\bibitem{Generalis:1983hb}
  S.~C.~Generalis and D.~J.~Broadhurst,
  Phys.\ Lett.\ B {\bf 139}, 85 (1984).

\bibitem{Mason:2005bj}
  Q.~Mason, H.~D.~Trottier, R.~Horgan, C.~T.~H.~Davies and G.~P.~Lepage
                  [HPQCD Collaboration],
  Phys.\ Rev.\ D {\bf 73}, 114501 (2006).

\bibitem{Narison:2002hk}
  S.~Narison,
  hep-ph/0202200.

\bibitem{Barate:1999hj}
  R.~Barate {\it et al.}  [ALEPH Collaboration],
  Eur.\ Phys.\ J.\ C {\bf 11}, 599 (1999).
  The spectral function in Fig.~9 of
  this reference is related to ${\rm Im}\Pi_1$ in (\ref{eq:Pi1}) by  
  $v_1^S(t) = 2\pi\,{\rm Im}\Pi_1(t)$.  

\bibitem{Lepage:1980fj}
  G.~P.~Lepage and S.~J.~Brodsky,
  Phys.\ Rev.\ D {\bf 22}, 2157 (1980).

\bibitem{Chetyrkin:1996sr}
  K.~G.~Chetyrkin,
  Phys.\ Lett.\ B {\bf 390}, 309 (1997).

\bibitem{Hill:2005ju}
  R.~J.~Hill,
  Phys.\ Rev.\ D {\bf 73}, 014012 (2006).

\bibitem{Gasser:1984ux}
  J.~Gasser and H.~Leutwyler,
  Nucl.\ Phys.\ B {\bf 250}, 517 (1985).

\bibitem{Jamin:2006tj}
  For some recent analysis along these lines, see: 
  M.~Jamin, J.~A.~Oller and A.~Pich,
  hep-ph/0605095.
  V.~Bernard, M.~Oertel, E.~Passemar and J.~Stern,
  hep-ph/0603202.

\bibitem{Bourrely:2005hp}
  C.~Bourrely and I.~Caprini,
  Nucl.\ Phys.\ B {\bf 722}, 149 (2005).

\end{thebibliography}
\end{document}